\documentclass[twocolumn,nofootinbib]{revtex4}

\usepackage{blindtext,hyperref,mathtools,amsmath,amsfonts,amssymb,physics,slashed,esvect,mathrsfs,dsfont,bbold}
\usepackage[utf8]{inputenc}
\usepackage{graphicx}
\usepackage{float}
\usepackage[footnotesize]{caption}

\usepackage{subfig}

\usepackage{multirow}

\usepackage{empheq}

\usepackage{pstricks,pst-node}

\DeclareCaptionJustification{justified}{\leftskip=0pt \rightskip=0pt \parfillskip=0pt plus 1fil}

\usepackage{xcolor}

\mathchardef\Re="023C 
\mathchardef\Im="023D

\pdfoutput=1

\usepackage{environ}
\NewEnviron{myequation}{
\begin{small}
\begin{equation}
\BODY
\end{equation}
\end{small}
}

\begin{document}
\title{Piecewise polytropic meshing and refinement method for the reconstruction of the neutron star equation of state using tidal deformabilities   and constraints in the piecewise polytropic parameters given by the GW170817 event}

\author{Juan Mena Fernández}
\email[]{juanmena@ucm.es}
\author{Luis Manuel González-Romero}
\email[]{mgromero@fis.ucm.es}
\affiliation{Departamento de Física Teórica, Facultad de Ciencias Físicas, Universidad Complutense de Madrid, 28040 Madrid, Spain}

\date{\today}

\begin{abstract}
In this paper we present a new approach to the inverse problem for relativistic stars using the piecewise polytropic parametrization of the equation of state. The algorithm is a piecewise polytropic meshing and refinement method that reconstructs the neutron star equation of state from experimental data of the mass and the tidal Love parameter. We use an initial mesh of $65536$ equations of state in a $4$-volume of piecewise polytropic parameters that contains most of the candidate equations of state used today. The refinement process drives us to the reconstruction of the equation of state with a certain precision. Using the reconstructed  equation of state, we calculate predictions for quasinormal modes and slow rotation parameters.

In order to check the meshing and refinement method, we use as input data a few (6) configurations of a given equation of state. We reconstruct the equation of state in a quite good approximation, and then we compare the curves of physical parameters from the original equation of state and the reconstructed one. We obtain a relative difference for all the parameters smaller than $7.5\%$.

We also study the constraints that impose the GW170817 event on the piecewise polytropic parameters $\{\log_{10}p_1,\Gamma_1,\Gamma_2,\Gamma_3\}$. We use the waveform model TaylorF2 for the low-spin scenario, and see that the EOSs that lie outside the $90\%$ credible region when $\bar{\lambda}^{tid}_1=\bar{\lambda}^{tid}_2$ define a zone of polytropic parameters that does not depend on $\Gamma_3$.
\end{abstract}

\pacs{1}

\maketitle

\section{Introduction}

The detection of gravitational waves by the LIGO-VIRGO collaboration (GW150914-\cite{abbott2016observation}, GW170814-\cite{abbott2017gw170814}, GW170817-\cite{abbott2017gw170817}) opens a new era in relativistic astrophysics. In particular, the GW170817 event, which seems to be the consequence of the merging and colliding of a pair of neutron stars, can be used to study the properties of these relativistic stars. In fact, binary mergers containing at least one neutron star offer a new possibility to constraint the EOS of matter at supranuclear densities. 

In principle, several observations of isolated and binary neutron stars may provide a set of pairs of mass-radius $(M,R)$ or mass-tidal Love parameter $(M,\bar{\lambda}^{tid})$ dense and accurate enough to reconstruct the neutron star EOS. The problem to obtain the EOS for neutron stars from macroscopic data of these stars has been treated by Lindblom using the mass-radio curve in \cite{lindblom1992determining}, recent modifications can be found in (\cite{lindblom2012spectral},\cite{lindblom2014spectral},\cite{lindblom2014relativistic}). This problem receives generally the name of inverse stellar structure problem. Other authors have studied the inverse problem using different techniques (\cite{kokkotas2001inverse},\cite{lackey2015reconstructing},\cite{carney2018comparing},\cite{abdelsalhin2018solving},\cite{abbott2018gw170817},\cite{volkel2019inverse}). In this paper we develop a method to reconstruct the EOS of neutron stars from a collection of pairs $(M,\bar{\lambda}^{tid})$. Similar problems have been treated by several authors since the observation of the GW170817 event (\cite{raithel2018tidal},\cite{chatziioannou2018measuring},\cite{most2018new},\cite{de2018tidal}). Here we propose a new approach to the inverse problem based in the piecewise polytropic meshing and refinement method presented in a recent paper \cite{mena2019reconstruction}, where we used w-quasinormal modes (QNMs) spectra to reconstruct the equation of state.

Our method to solve the inverse problem requires to generate a wide mesh of EOSs in a $4$-volume of piecewise polytropic parameters (we will generate a total of $65536$ EOSs in the initial mesh). Since it will be necessary to calculate the $\bar{\lambda}^{tid}(M)$ curve for each EOSs in the mesh, we will take advantage of these curves to study the restrictions that impose the GW170817 event to the piecewise polytropic parameters.

In section \ref{overview} we briefly summarize the necessary theoretical background, starting with static and spherically symmetric stars in order to introduce tidal deformations. In section \ref{code} we verify that the programs developed for the calculation of the tidal Love parameter work correctly using different types of EOSs. In section \ref{problema_inverso} we develop our piecewise polytropic meshing and refinement method to solve the inverse problem. In section \ref{problema_inverso_resultados} we test the method with an explicit example by using $6$ APR4 configurations as input data. In section \ref{constraintsGW170817} we study the constraints that impose the GW170817 event on the piecewise polytropic parameters, and also together with the $2M_\odot$ constraint. Finally, in section \ref{conclusions} we finish the paper with a summary of the main results.

\section{Overview of the formalism}\label{overview}

Here we will show the necessary differential equations to calculate the tidal Love parameter of non-rotating neutron stars. We will start with static and spherically symmetric relativistic stars and then we will introduce tidal perturbations.

\subsection{Static and spherically symmetric relativistic stars}

Coordinates can be chosen so that the line element has the form
\begin{myequation}\label{sim_esf}
	ds^2=-e^{\nu(r)}(cdt)^2+e^{\lambda(r)}dr^2+r^2(d\theta^2+\sin^2\theta d\varphi^2).
\end{myequation}
We will consider the matter in the interior of the star as an effective perfect fluid with a barotropic equation of state.
$u^\mu$ is the fluid's $4$-velocity, $p$ is the pressure and $\epsilon$ is the energy density$/c^2$. 

It is widely known that the equations describing static and spherically symmetric relativistic stars are given by
\begin{subequations}\label{orden0}
	\begin{gather}
	\begin{align}
	&\frac{dm}{dr}=4\pi r^2\epsilon,\\
	&\frac{d\nu}{dr}=\frac{2G}{c^2r^2}\frac{m+\frac{4\pi}{c^2}r^3p}{1-\frac{2Gm}{c^2r}},\\
	&\frac{dp}{dr}=-\left(\epsilon+\frac{p}{c^2}\right)\frac{G}{r^2}\frac{m+\frac{4\pi}{c^2}r^3p}{1-\frac{2Gm}{c^2r}},
	\end{align}
	\end{gather}
\end{subequations}
where
\begin{myequation}
	m=\frac{c^2r}{2G}(1-e^{-\lambda}).
\end{myequation}
Provided an equation of state, $p=p(\epsilon)$, the system of ordinary differential equations \eqref{orden0} can be solved numerically.

\subsection{Tidally deformed relativistic stars}

Restricting our analysis to the $l=2$ perturbations in the Regge-Wheeler gauge (\cite{regge1957stability},\cite{thorne1967non},\cite{hinderer2008tidal}), the full line element can be expressed as
\begin{myequation}
	\begin{array}{ll}
		ds^2=-H^2(r,\theta)(cdt)^2+Q^2(r,\theta)dr^2\\\\
		+r^2K^2(r,\theta)(d\theta^2 +\sin^2\theta d\varphi^2),
	\end{array}
\end{myequation}
where
\begin{small}
	\begin{subequations}
		\begin{gather}
		\begin{align}
		&H^2(r,\theta)=e^{\nu(r)}\left[1+2h_2(r)P_2(\cos\theta)\right],\\
		&Q^2(r,\theta)=e^{\lambda(r)}\left[1+\frac{2G}{c^2}\frac{e^{\lambda(r)}}{r}m_2(r)P_2(\cos\theta)\right],\\
		&K^2(r,\theta)=1+2k_2(r)P_2(\cos\theta).
		\end{align}
		\end{gather}
	\end{subequations}
\end{small}

Writing down the Einstein equations, one finds a first integral of motion,
\begin{myequation}
h_2+\frac{Gm_2}{c^2r}e^\lambda=0,
\end{myequation}
and a system of two ordinary differential equations,
\begin{small}
	\begin{subequations}
		\begin{gather}
		\begin{align}
		&\frac{dv}{dr}=-h_2\nu',\\
		&\frac{dh_2}{dr}=\left\{-\nu'+\frac{1}{c^2\left(1-\frac{2Gm}{rc^2}\right)\nu'}\left[8\pi G\left(\epsilon+\frac{p}{c^2}\right)\right.\right.\nonumber\\
		& \qquad -\left.\left.\frac{4Gm}{r^3}\right]\right\}h_2-\frac{4v}{r^2\left(1-\frac{2Gm}{rc^2}\right)\nu'},
		\end{align}
		\end{gather}
	\end{subequations}
\end{small}
where $v=h_2+k_2$. Once these equations have been numerically solved, we will calculate the tidal Love number, $k_2^{tid}$. The tidal Love number is related to how easy or difficult it would be to deform a star. It is given by \cite{hinderer2008tidal}
\begin{myequation}
	\begin{array}{ll}
		k_2^{tid}=\frac{8C^5}{5}(1-2C)^2\left[2+2C(y-1)-y\right]\left\{2C\left[6-3y\right.\right.\\\\
		\left.+3C(5y-8)\right]+4C^3\left[13-11y+C(3y-2)+2C^2(1+y)\right]\\\\
		\left.+3(1-2C)^2\left[2-y+2C(y-1)\right]\log(1-2C)\right\}^{-1},
	\end{array}
\end{myequation}
where
\begin{myequation}
	C=\frac{GM}{c^2R},\quad y=R\left[\frac{1}{h_2}\frac{dh_2}{dr}\right]\bigg\rvert_{r=R}-\frac{4\pi R^3\epsilon_{sup}}{M}.
\end{myequation}
$C$ is known as the compactness parameter, and $\epsilon_{sup}$ is the energy density$/c^2$ at the surface of the star, if non-zero \cite{hinderer2010tidal}. We will be interested in calculating the so-called tidal Love parameter, which is given in terms of the tidal Love number $k_2^{tid}$ and the compactness parameter $C$ as
\begin{myequation}
	\bar{\lambda}^{tid}=\frac{2k_2^{tid}}{3C^5}.
\end{myequation}

\section{The code analysis}\label{code}

In order to check the codes developed to solve the equations obtained in section \ref{overview}, we have used well known equations of state of different types (EOS with plain nuclear matter, with hyperons, for hybrids stars and for quark stars). We list below the different models of EOSs considered in this paper.

\begin{itemize}
	\item For plain $npe\mu$ nuclear matter we use
	\begin{itemize}
		\item APR4 EOS \cite{akmal1998equation}, obtained using a variational method.
		\item SLy EOS \cite{douchin2001unified}, obtained using a potential-method.
	\end{itemize}
	\item For mixed hyperon-nuclear mater we use
	\begin{itemize}
		\item GNH3 EOS \cite{glendenning1984neutron}, a relativistic mean-field theory EOS containing hyperons.
		\item BHZBM EOS \cite{bednarek2012hyperons}, a non-linear relativistic mean field model involving baryon octet coupled to meson fields.
	\end{itemize}
	\item For hybrid stars we use ALF4 EOS \cite{alford2005hybrid}, a hybrid EOS with mixed APR nuclear matter and color-flavor-locked quark matter.
	\item For hybrid stars with hyperons and quark color-superconductivity we use BS3 EOS \cite{bonanno2012composition}, obtained using a combination of phenomenological relativistic hyper-nuclear density functional and an effective NJL model of quantum chromodynamics. The parameters considered are vector coupling $G_V/G_S=0.6$ and quark-hadron transition density $\rho_{tr}/\rho_0=3.5$, where $\rho_0$ is the density of nuclear saturation.
	\item For quark stars we use WSPHS EOS \cite{weissenborn2011quark}, an unpaired quark matter EOS with parameters $B_{eff}^{1/4}=123.7$ MeV and $a_4=0.53$.
\end{itemize}

\begin{figure}[H]
\centering
\includegraphics[width=\linewidth]{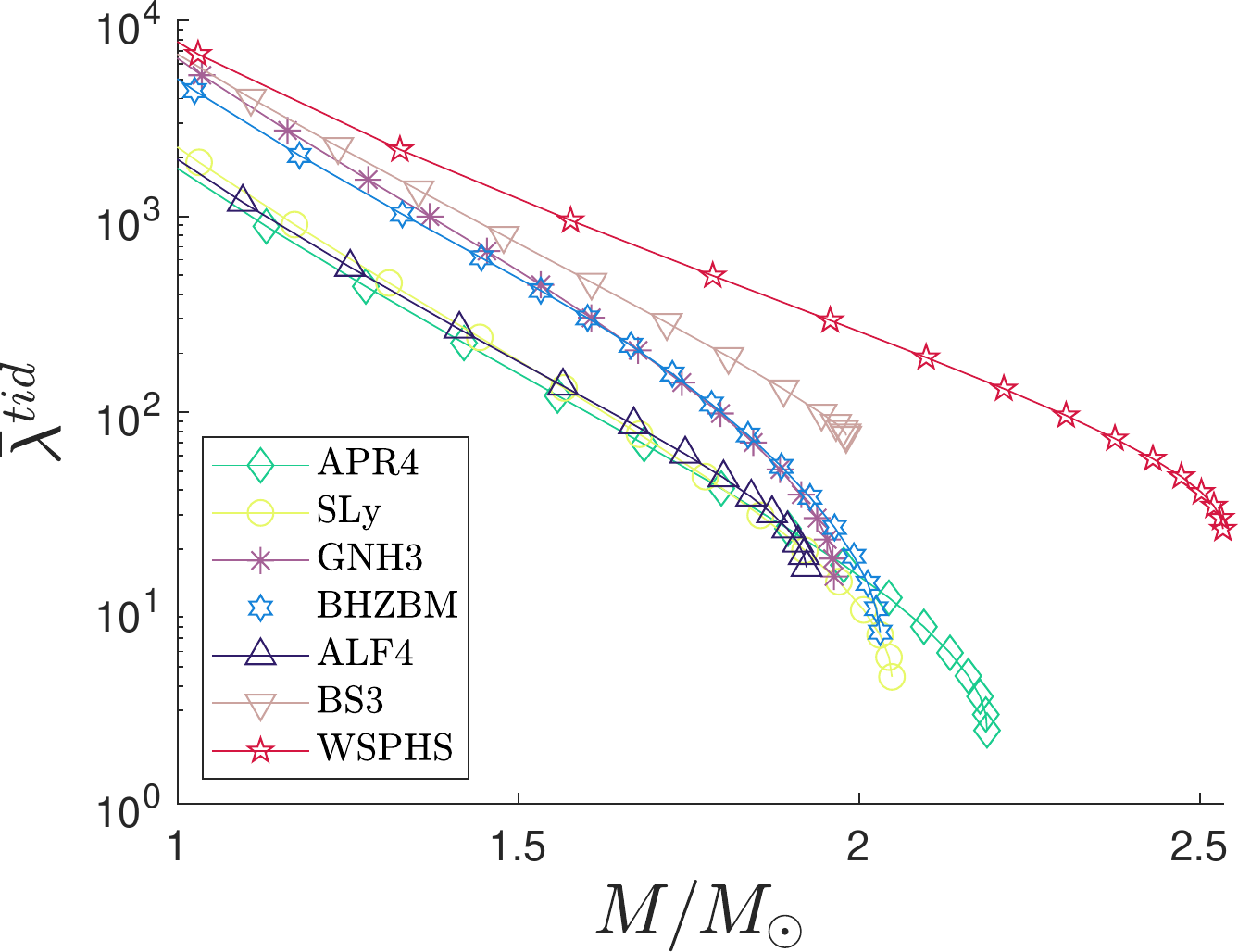}\captionsetup{width=0.5\textwidth}\caption{Tidal Love parameter vs mass for the different EOSs considered.}\label{fig2}
\end{figure}

The results of applying the codes to these models can be found in FIG. \ref{fig2}. In this figure we present the tidal Love parameter versus the mass ($M-\bar{\lambda}^{tid}$ curve) for the different EOSs considered, with $20$ configurations for each EOS. 

For our method to reconstruct the EOS of neutron stars, it will be necessary to calculate thousands of these $M-\bar{\lambda}^{tid}$ curves, as will be explained in section \ref{problema_inverso}. Once the EOS is reconstructed, we will be able to make predictions of other macroscopic parameters. We will consider  parameters calculated for slowly rotating relativistic stars, together with the axial quasinormal modes. The necessary equations and algorithms we will use to calculate these parameters can be found in reference \cite{mena2019reconstruction}.

In section \ref{constraintsGW170817} we will study the constraints that impose the GW170817 event for thousands of EOSs generated with the piecewise polytropic parametrization. Now we will study these constraints for the EOSs considered in this section. 

The observation of gravitational waves provides new information about which models of EOSs are more likely to be candidate EOSs. The GW170817 event was the first observation of gravitational waves from a binary neutron star inspiral \cite{abbott2017gw170817}. Consider the waveform model TaylorF2 for the GW170817 event of reference \cite{abbott2019properties}. This model leads to a chirp mass given by

\begin{myequation}\label{chirp_mass}
\mathcal{M}=(1.186 \pm 0.001) M_\odot
\end{myequation}
and a mass ratio given by
\begin{myequation}\label{mass_ratio}
\frac{M_2}{M_1}=[0.72,1]
\end{myequation}
for the low-spin scenario. From now on, we will only consider the low-spin scenario because it matches observations of binary neutron stars in our Galaxy \cite{tauris2017formation}. 

\begin{figure}[H]
\centering
\includegraphics[width=\linewidth]{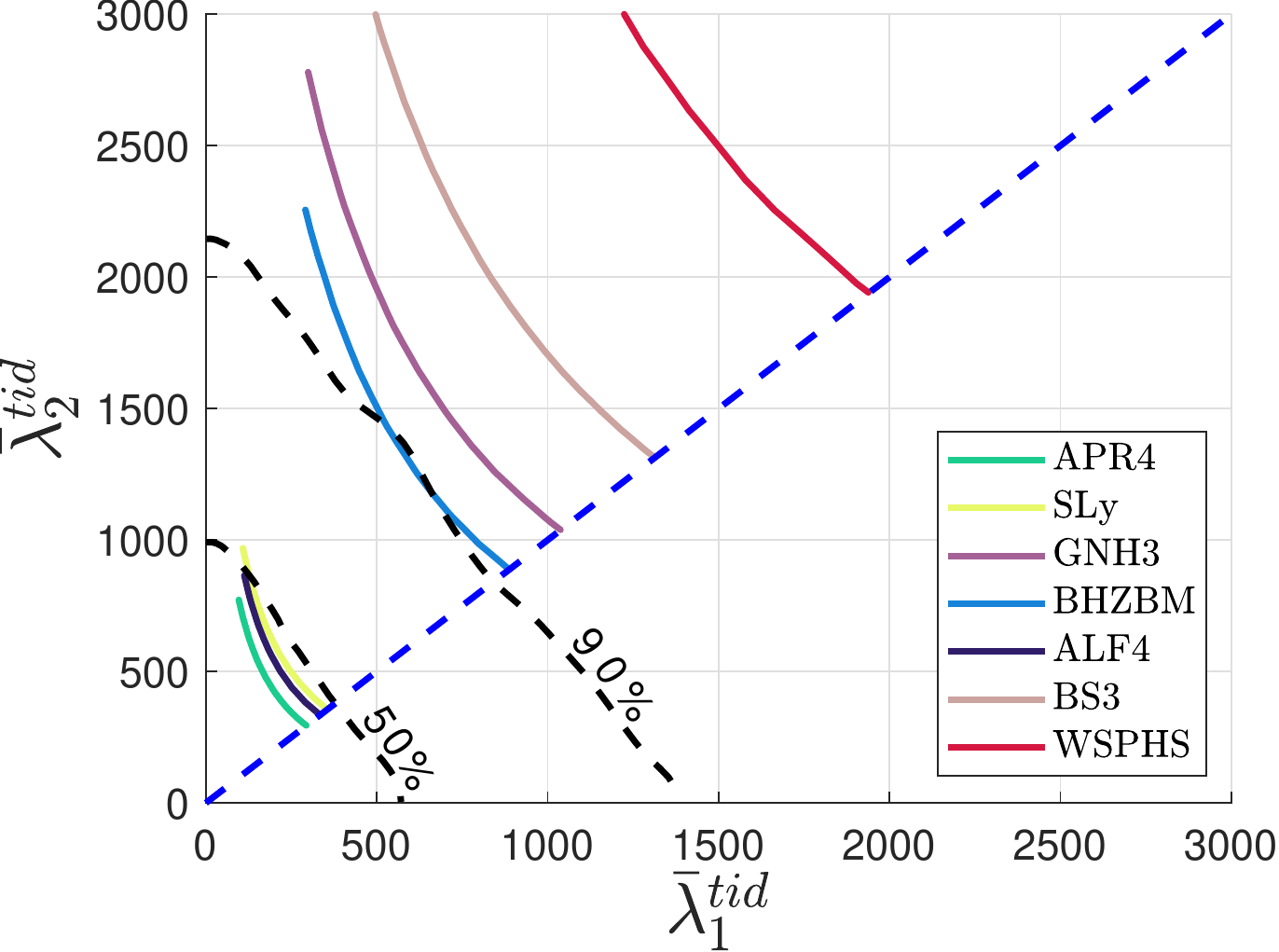}
\captionsetup{width=0.5\textwidth}\caption{Predictions for tidal deformability given by the different realistic EOSs considered in this paper, under the assumption that both components are neutron stars. Contours enclosing $90\%$ and $50\%$ of the probability density are shown as dashed lines (both curves taken from reference \cite{abbott2019properties}).}\label{fig3}
\end{figure}

With eqs. \eqref{chirp_mass} and \eqref{mass_ratio} one could calculate the  possible values of the masses of both stars by using the definition of the chirp mass,
\begin{myequation}\label{chirp_mass_equation}
\mathcal{M}=\frac{(M_1M_2)^{3/5}}{(M_1+M_2)^{1/5}}.
\end{myequation}
Once this is done, one could also calculate the corresponding possible values of the tidal Love parameters of both stars for a given EOS. The resulting curves for the different EOSs considered in this paper are shown in FIG. \ref{fig3}, together with the contours enclosing $90\%$ and $50\%$ of the probability density (curves taken from FIG. 10 of reference \cite{abbott2019properties}). The lengths of these curves are determined by the uncertainty in the mass ratio, eq. \eqref{mass_ratio}.

In FIG. \ref{fig3} we observe that GNH3, BHZBM, BS3 and WSPHS EOSs predict $\bar{\lambda}^{tid}$ values outside the $90\%$ credible region.

\section{The piecewise polytropic meshing and refinement method for the inverse problem}\label{problema_inverso}

The meshing and refinement method explained in this section is analogous to the one we developed in reference \cite{mena2019reconstruction}. The main difference is that here the algorithm starts with a set of pairs of mass-tidal Love parameter $(M,\bar{\lambda}^{tid})$ instead of mass-frequency of the fundamental wI mode $(M,\nu)$.

The complete knowledge of the neutron star EOS makes possible the calculation of macroscopic quantities such as the mass, the quasinormal modes, the tidal Love parameter, etc. Viceversa, from the measurement of macroscopic observables it is possible to invert this map and reconstruct the EOS: this is the so-called \textit{inverse problem} \citep{lindblom1992determining}.

\begin{figure}[H]
	\centering
	\includegraphics[width=\linewidth]{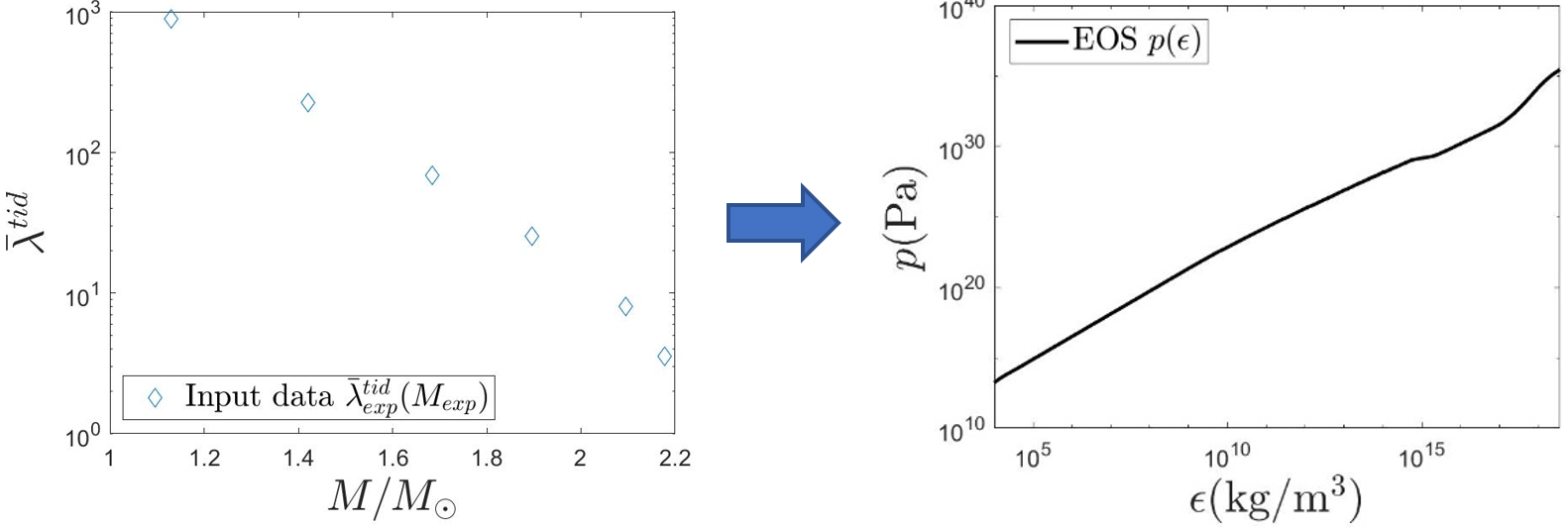}\captionsetup{width=0.5\textwidth}\caption{Illustration of the inverse problem for neutron stars. The EOS is reconstructed from measurements of the tidal Love parameter and the mass of $6$ different neutron stars.}\label{fig5}
\end{figure}

The piecewise polytropic parametrization fits a large class of realistic and candidate EOSs \cite{read2009constraints}. In fact, the polytropic parameters $\{\log_{10}p_1,\Gamma_1,\Gamma_2,\Gamma_3\}$ for a wide variety of EOSs can be found in TABLE III of reference \cite{read2009constraints}. Thus, each EOS is determined simply by specifying 4 numbers. Here is where the idea of our inverse stellar method arises: we will create a mesh of EOSs in a $4$-volume of piecewise polytropic parameters $\{\log_{10}p_1,\Gamma_1,\Gamma_2,\Gamma_3\}$. This mesh has to include as many candidate EOSs as possible, for example, from the ones listed in TABLE III of reference \cite{read2009constraints}. The initial mesh of polytropic parameters we chose is given by 
\begin{myequation}\label{parametrosI}
	\begin{array}{ll}
		\log_{10}p_1=[34,34.7]_{16},\\\\
		\Gamma_1=[2,4.1]_{16},\\\\
		\Gamma_2=[1.8,3.8]_{16},\\\\
		\Gamma_3=[1.8,3.8]_{16}.
	\end{array}
\end{myequation}
From now on, the sub-index in an interval will indicate the number of equidistant elements taken in that interval. Hence, we will have a total of $16^4=65536$ EOSs in our initial $4$-volume, i.e. $65536$ points in a $4$-dimensional space of coordinates $\{\log_{10}p_1,\Gamma_1,\Gamma_2,\Gamma_3\}$.

As shown in the illustration of the inverse problem, FIG. \ref{fig5}, we will reconstruct the neutron star EOS from measurements of the tidal Love parameter ($\bar{\lambda}^{tid}$) and the mass ($M$) of some different neutron stars. The input data will be denoted as $M_{\text{exp}}$ and $\bar{\lambda}^{tid}_{\text{exp}}$.

Our algorithm will numerically calculate each $\bar{\lambda}^{tid}_i(M_i)$ curve ($i=1,\dots,65536$) in order to find the most similar $\bar{\lambda}^{tid}(M)$ curve to the input data $\bar{\lambda}^{tid}_{\text{exp}}(M_{\text{exp}})$. That is, it will find the point in the $4$-space of coordinates $\{\log_{10}p_1,\Gamma_1,\Gamma_2,\Gamma_3\}$ that represents the input data with a certain precision.

\begin{figure}[H]
\centering
\includegraphics[width=0.53\linewidth]{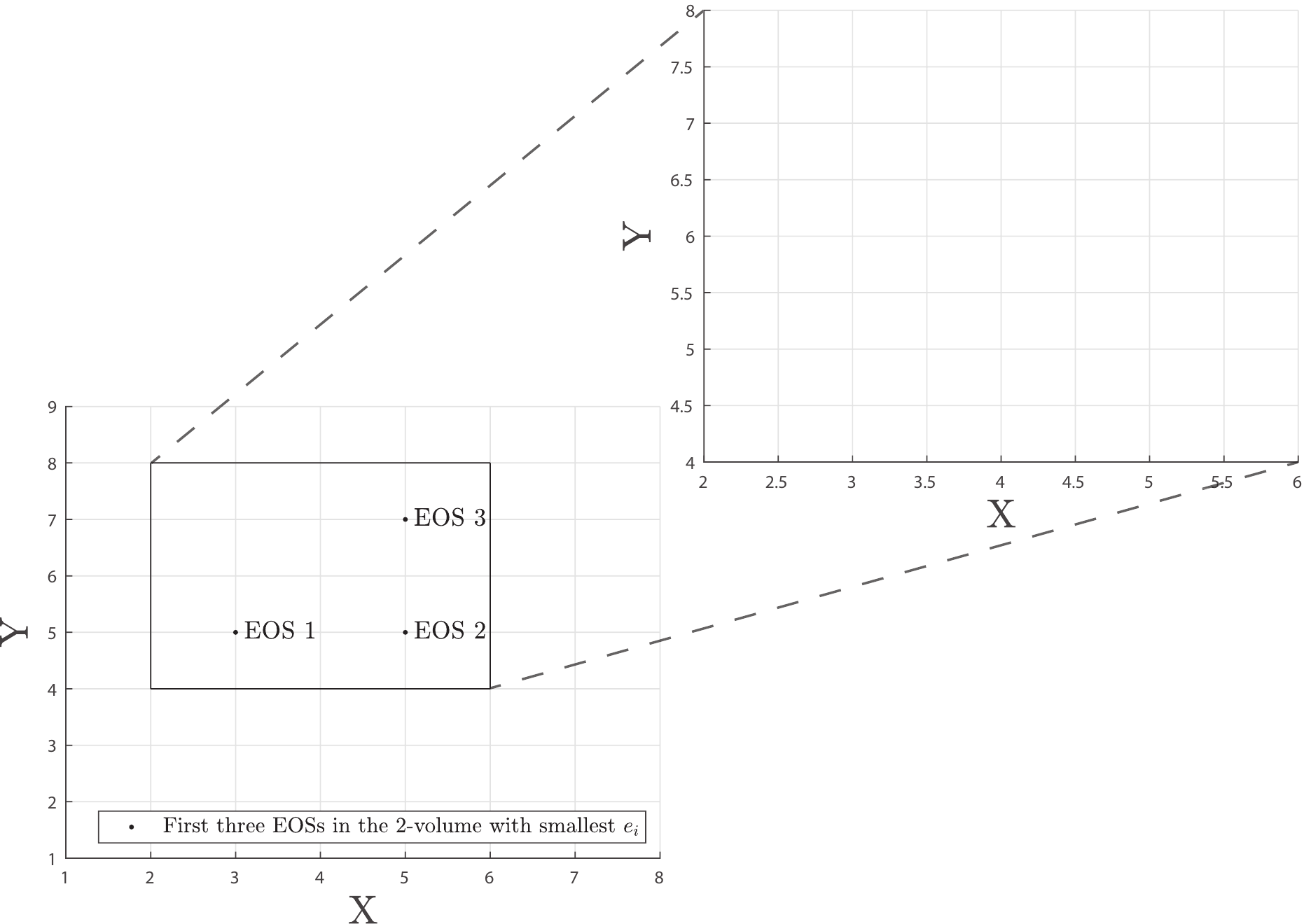}\captionsetup{width=\textwidth}\caption{Graphical illustration of the local refinement of the initial mesh from the three EOSs with smallest values of $e_i$. The image represents a simplified model with $2$ polytropic parameters $X=[1,8]_8$ and $Y=[1,9]_9$.}\label{fig6}
\end{figure}

The algorithm proceeds as follows:
\begin{enumerate}
	\item calculate the tidal Love parameter ($\bar{\lambda}^{tid}$) and the mass ($M$) for every EOS in the $4$-volume of polytropic parameters. We calculate $30$ configurations for each EOS in the same fixed central pressure range, namely $\log_{10}p_0=[33.9,36.714]_{30}$.
	\item fit with piecewise linear interpolation the curve $\bar{\lambda}^{tid}_i(M_i)$ for all the EOSs in the $4$-space of polytropic parameters. The interpolation is necessary to calculate $\bar{\lambda}^{tid}_i(M_{\text{exp}})$.
	\item compare each curve $\bar{\lambda}^{tid}_i(M_i)$ with the input data $\bar{\lambda}^{tid}_{\text{exp}}(M_{\text{exp}})$ by calculating
	\begin{myequation}
		e_i=\max\left\{\frac{\abs{\bar{\lambda}^{tid}_i(M_{\text{exp}})-\bar{\lambda}^{tid}_{\text{exp}}(M_{\text{exp}})}}{\bar{\lambda}^{tid}_{\text{exp}}(M_{\text{exp}})}\right\}.
	\end{myequation}
	
	We only calculate $e_i$ for those EOSs that fulfill the condition $\max (M_i)\geq\max (M_{\text{exp}})$. The smaller $e_i$ is, the more similar $\bar{\lambda}^{tid}_i(M_i)$ and $\bar{\lambda}^{tid}_{\text{exp}}(M_{\text{exp}})$ are.
	\item sort the EOSs in increasing order of $e_i$ and check the value of $\min_i(e_i)$.
	\item if $\min_i(e_i)<\text{tol}$, finish the algorithm. In other case, define a new $4$-volume of polytropic parameters that contains, for example, the first $3$ EOSs with smallest $e_i$, and repeat from step $1$. This new $4$-volume of polytropic parameters is a local refinement of the initial mesh. A graphical illustration of the local refinement is shown in FIG. \ref{fig6}.
\end{enumerate}

\onecolumngrid

\twocolumngrid

\begin{figure}[H]
\centering
\includegraphics[width=\linewidth]{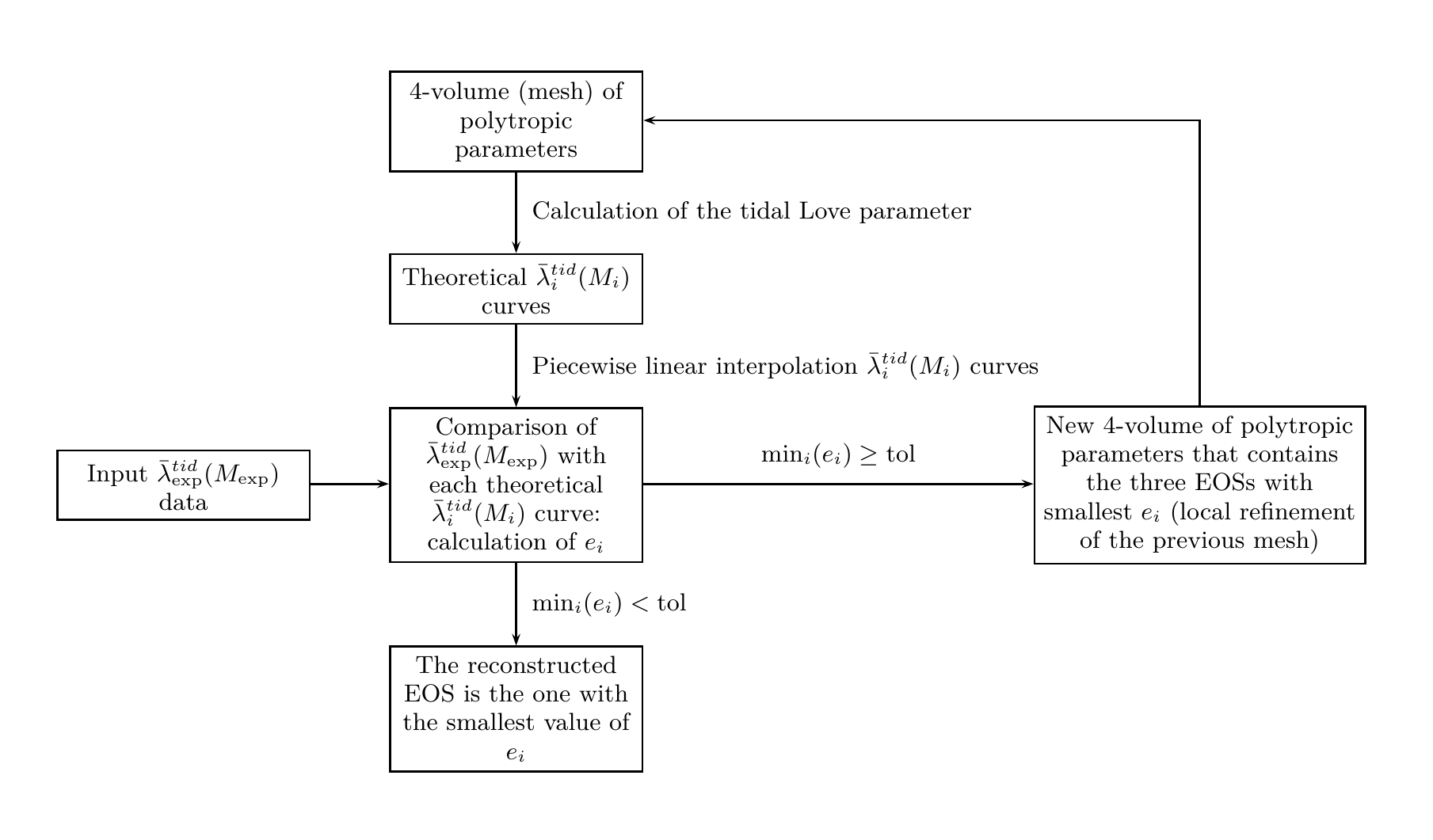}\captionsetup{width=\textwidth}\caption{Scheme of the piecewise polytropic meshing and refinement method for the inverse problem.}\label{fig7}
\end{figure}

A scheme of the entire meshing and refinement method is shown in FIG. \ref{fig7}.

\onecolumngrid

\twocolumngrid

Once the algorithm has finished, we will have a reconstructed EOS. Then, we will be able to calculate other macroscopic parameters (such as the moment of inertia, the QNMs,\dots) in order to make predictions that can be compared with other experimental data.

Since today we still do not have the necessary experimental data, in order to test the algorithm we will suppose that the measured macroscopic parameters correspond to, for example, the ones calculated for APR4 EOS ($M_{\text{exp}}=M_{\text{APR4}}$ and $\bar{\lambda}^{tid}_{\text{exp}}=\bar{\lambda}^{tid}_{\text{APR4}}$). We will consider $6$ of the $20$ APR4 configurations shown in FIG. \ref{fig2}. Since APR4 is a known EOS, we will be able to directly compare the reconstructed EOS with the original one and also to compare them in macroscopic parameters.

The numerical results of the meshing and refinement method are described in section \ref{problema_inverso_resultados}.

\section{Testing the meshing and refinement method}\label{problema_inverso_resultados}

Here we will consider 6 APR4 EOS stellar configurations as our input data, i.e. $M_{\text{exp}}=M_{\text{APR4}}$ and $\bar{\lambda}^{tid}_{\text{exp}}=\bar{\lambda}^{tid}_{\text{APR4}}$, in order to test the meshing and refinement method explained in section \ref{problema_inverso}. To carry out the test we will consider a tolerance $\text{tol}=0.02$.

\vspace{0.07cm}\begin{center} \textbf{First iteration of the method}\end{center}\vspace{0.07cm}

We proceed as explained in section \ref{problema_inverso} (steps 1. to 4.). The polytropic parameters of the first 3 EOSs with the smallest values of $e_i$ in the mesh of EOSs given by eq. \eqref{parametrosI} are listed in TABLE \ref{tabla1}.

\begin{table}[H]
	\begin{center}
		\begin{tabular}{| c | c | c | c | c |}\hline
			$e_i$ & $\log_{10} p_1$ & $\Gamma_1$ & $\Gamma_2$ & $\Gamma_3$\\\hline
0.043519 & 34.28 & 2.84 & 3.4 & 3.4\\
0.044324 & 34.233 & 2.84 & 3.5333 & 3.4\\
0.046073 & 34.28 & 2.7 & 3.4 & 3.4\\\hline
		\end{tabular}
	\end{center}
	\captionsetup{width=0.5\textwidth}\caption{Polytropic parameters of the first 3 EOSs with the smallest values of $e_i$ in the $4$-volume defined by eq. \eqref{parametrosI}.}\label{tabla1}
\end{table}

Since $\min_i(e_i)\geq 0.02$, we proceed with the refinement of the initial mesh (step 5. of our method). Hence, we define a local refinement of the initial mesh of piecewise polytropic parameters that contains the EOSs listed in TABLE \ref{tabla1}. We will explain the refinement process with an explicit example. The original $\Gamma_2$ vector, eq. \eqref{parametrosI}, was given by
\begin{myequation}
\Gamma_2=[1.8,3.8]_{16}.
\end{myequation}
The difference between two elements in this vector is given by
\begin{myequation}
\delta\Gamma_2=\frac{3.8-1.8}{16-1}=0.1333.
\end{myequation}
We define the $\Gamma_2$ vector for the next iteration as
\begin{myequation}
\Gamma_2^{\text{new}}=[\min\Gamma_2(\text{TABLE I})-\delta\Gamma_2,\max\Gamma_2(\text{TABLE I})+\delta\Gamma_2].
\end{myequation}
By taking a look at TABLE \ref{tabla1}, we find that the refinement of $\Gamma_2$ is, then, given by
\begin{myequation}
\Gamma_2^{\text{new}}=[3.2667,3.6667].
\end{myequation}
This refinement process allows our method to determine the EOS even if its polytropic parameters do not belong to the original $4$-volume. We proceed analogously with the other polytropic parameters and find that the new mesh of EOSs is given by
\begin{myequation}\label{parametrosII}
	\begin{array}{ll}
		\log_{10}p_1=[34.1867,34.3267]_{10},\\\\
		\Gamma_1=[2.56,2.98]_{10},\\\\
		\Gamma_2=[3.2667,3.6667]_{10},\\\\
		\Gamma_3=[3.2667,3.5333]_{10}.
	\end{array}
\end{myequation}
We have chosen each vector to have a total of $10$ elements.

\vspace{0.07cm}\begin{center} \textbf{Second iteration of the method (first refinement)}\end{center}\vspace{0.07cm}

The polytropic parameters of the first 3 EOSs with the smallest values of $e_i$ in the mesh of EOSs given by eq. \eqref{parametrosII} are listed in TABLE \ref{tabla2}.

\begin{table}[H]
	\begin{center}
		\begin{tabular}{| c | c | c | c | c |}\hline
			$e_i$ & $\log_{10} p_1$ & $\Gamma_1$ & $\Gamma_2$ & $\Gamma_3$\\\hline
0.021242 & 34.265 & 2.8867 & 3.4445 & 3.3852\\
0.021384 & 34.265 & 2.9333 & 3.4445 & 3.3852\\
0.021632 & 34.249 & 2.7467 & 3.4889 & 3.4148\\\hline
		\end{tabular}
	\end{center}
	\captionsetup{width=0.5\textwidth}\caption{Polytropic parameters of the first 3 EOSs with the smallest values of $e_i$ in the $4$-volume defined by eq. \eqref{parametrosII}.}\label{tabla2}
\end{table}

Since $\min_i(e_i)\geq 0.02$, we proceed with the refinement process (step 5. of our method). The local refinement of the mesh of EOSs given by eq. \eqref{parametrosII} is given by

\begin{myequation}\label{parametrosIII}
	\begin{array}{ll}
		\log_{10}p_1=[34.2337,34.2803]_{10},\\\\
		\Gamma_1=[2.7,2.98]_{10},\\\\
		\Gamma_2=[3.4,3.5334]_{10},\\\\
		\Gamma_3=[3.3556,3.4444]_{10}.
	\end{array}
\end{myequation}
In the third iteration we will have a total of $10^4=10000$ EOSs. 

\vspace{0.07cm}\begin{center} \textbf{Third iteration of the method (second refinement)}\end{center}\vspace{0.07cm}

The polytropic parameters of the first 3 EOSs with the smallest values of $e_i$ in the mesh of EOSs given by eq. \eqref{parametrosIII} are listed in TABLE \ref{tabla3}.

\begin{table}[H]
	\begin{center}
		\begin{tabular}{| c | c | c | c | c |}\hline
			$e_i$ & $\log_{10} p_1$ & $\Gamma_1$ & $\Gamma_2$ & $\Gamma_3$\\\hline
0.014192 & 34.26 & 2.8556 & 3.4593 & 3.4049\\
0.014233 & 34.26 & 2.8556 & 3.4593 & 3.3951\\
0.014606 & 34.26 & 2.8244 & 3.4593 & 3.4049\\\hline
		\end{tabular}
	\end{center}
	\captionsetup{width=0.5\textwidth}\caption{Polytropic parameters of the first 3 EOSs with the smallest values of $e_i$ in the $4$-volume defined by eq. \eqref{parametrosIII}.}\label{tabla3}
\end{table}

Since $\min_i(e_i)<0.02$, we stop the algorithm. From now on, the first EOS listed in TABLE \ref{tabla3} will be denoted as the reconstructed APR4 EOS.

\vspace{0.07cm}\begin{center} \textbf{Comparison between the original and the reconstructed APR4 equations of state}\end{center}\vspace{0.07cm}

Here we will distinguish between three different APR4 EOS:
\begin{itemize}
	\item The original APR4 EOS \cite{akmal1998equation}.
	\item The reconstructed APR4 EOS. This is the one our algorithm reconstructed, whose polytropic parameters are listed in the first row of TABLE \ref{tabla3}.
	\item The polytropic APR4 EOS. This one is the polytropic fit of APR4 EOS, whose polytropic parameters can be found in TABLE III of reference \cite{read2009constraints}.
\end{itemize}

\begin{figure}[H]
	\centering
	\includegraphics[width=\linewidth]{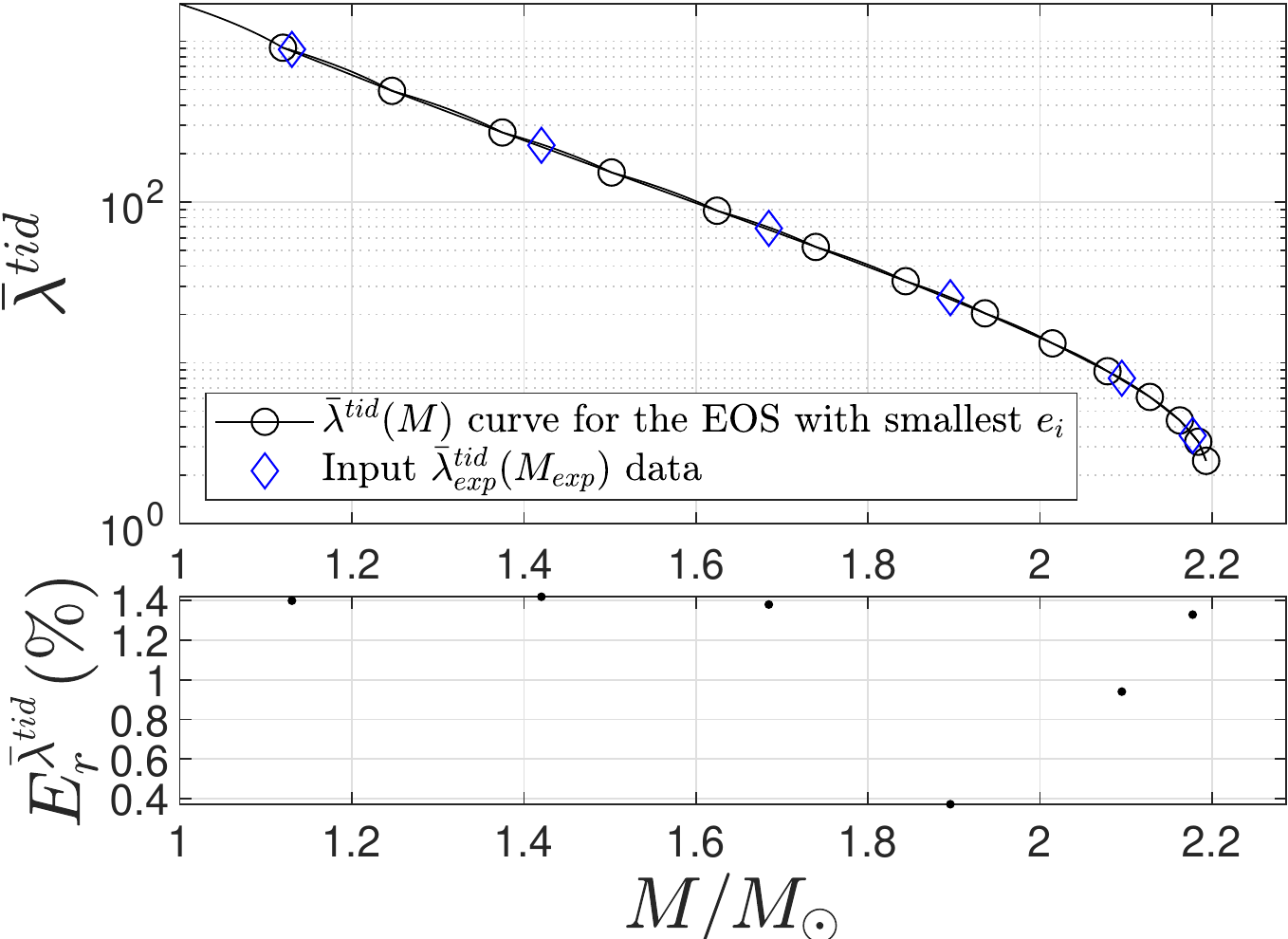}
	\captionsetup{width=0.5\textwidth}\caption{Top panel: tidal Love parameter vs mass for the input data (blue diamonds) and the reconstructed EOS (black circles). Bottom panel: relative difference between the input data and the reconstructed EOS.}\label{fig17}
\end{figure}

FIG. \ref{fig17} shows the tidal Love parameter vs the mass for the original APR4 EOS (blue diamonds) and for the reconstructed APR4 EOS (black circles), together with the relative difference.

Since the input data correspond to $6$ APR4 configurations, we can compare the reconstructed APR4 EOS with the original one (this would not be possible if we had used real experimental data). There are two different ways to compare them: 
\begin{enumerate}
\item Directly comparing the EOSs. The easiest way to do it is by comparing the polytropic parameters of the reconstructed EOS with those of the polytropic one. The polytropic parameters of both EOSs are listed in TABLE \ref{tabla4}.
	
\begin{table}[H]
\begin{center}
\begin{tabular}{| c | c | c | c | c |}\hline
EOS & $\log_{10} p_1$ & $\Gamma_1$ & $\Gamma_2$ & $\Gamma_3$\\\hline
Reconstructed APR4 & 34.26 & 2.8556 & 3.4593 & 3.4049\\
Polytropic APR4 & 34.269 & 2.830 & 3.445 & 3.348\\\hline
\end{tabular}
\end{center}
\captionsetup{width=0.45\textwidth}\caption{Comparison between the polytropic parameters of the reconstructed APR4 EOS and the polytropic one.}\label{tabla4}
\end{table}
	
The polytropic parameters of the reconstructed APR4 EOS are very similar to those of the polytropic one.
	
\item Comparing both EOSs in macroscopic parameters. In TABLE \ref{tabla5} we compare the reconstructed APR4 EOS with the original one by calculating $20$ stellar configurations with the same central energy densities for both EOSs. We also compare the original APR4 EOS with the polytropic one, in order to check how good our reconstruction is.
	
\begin{table}[H]
\begin{center}
\begin{tabular}{| c | c | c |}\hline
\multirow{ 2}{*}{Parameter} & \multicolumn{2}{c|}{Maximum relative difference ($\%$)}\\\cline{2-3}
& a) Original-reconstructed & b) Original-polytropic\\\hline
$p_0$ & $3.8167$ & $1.4693$\\
$R$ & $0.73509$ & $0.25324$\\
$M$ & $0.93748$ & $0.33116$\\
$I$ & $2.2526$ & $0.59308$\\
$\bar{I}$ & $0.69243$ & $0.90994$\\
$Q$ & $1.0479$ & $0.80456$\\
$\bar{Q}$ & $1.2427$ & $0.87492$\\
$\bar{\lambda}^{tid}$ & $6.3168$ & $2.0358$\\
$\nu$ & $0.66183$ & $0.134$\\
$\tau$ & $5.9958$ & $1.884$\\
$\Re \bar{\omega}$ & $2.505$ & $0.80374$\\
$\Im \bar{\omega}$ & $7.4071$ & $2.5624$\\
\hline
\end{tabular}
\end{center}
\captionsetup{width=0.5\textwidth}\caption{a) Maximum relative difference in macroscopic parameters between the original APR4 EOS and the reconstructed one out of $20$ configurations calculated. b) Maximum relative difference in macroscopic parameters between the original APR4 EOS and the polytropic one out of $20$ configurations calculated. We show the maximum relative difference in central pressure ($p_0$), radius ($R$), mass ($M$), moment of inertia ($I$), quadrupole moment ($Q$), I-Love-Q parameters ($\bar{I}$, $\bar{Q}$, $\bar{\lambda}^{tid}$), frequency and damping time of the fundamental wI mode ($\nu$ and $\tau$) and re-scaled $\omega$ of the fundamental wI mode ($\bar{\omega}$) (see reference \cite{mena2019reconstruction} for further details about the calculation of these parameters).}\label{tabla5}
\end{table}

The reconstructed APR4 EOS is very similar to the original one in macroscopic parameters. Comparing the results of both columns of TABLE \ref{tabla5}, we conclude that our polytropic reconstruction of APR4 EOS is quite good.

\end{enumerate}

We conclude that our reconstruction of APR4 EOS with the meshing and refinement method is very similar to the piecewise polytropic fit. This means that, starting from only $6$ input $(M,\bar{\lambda}^{tid})$ points, we have been able to reconstruct the neutron star EOS in a good approximation.

\subsection{Testing the meshing and refinement method with experimental error}

Consider the same $6$ APR4 configurations we used as input data for the meshing and refinement method. If the input data was experimental data, it would have an associated experimental error. In order to make an estimation of how this experimental error would affect the final results, now we will randomly modify the input data in a uncertainty interval, i.e.
\begin{myequation}\label{random_input}
X^i_{\text{exp}}\to X^i_{\text{exp}}+\text{rand}\left[-\Delta X^i_{\text{exp}},\Delta X^i_{\text{exp}}\right],\ i=1,\dots,5,
\end{myequation}
with 
\begin{myequation}
\Delta X_{\text{exp}}^i=\epsilon_X X_{\text{exp}}^i,\ i=1,\dots,5,
\end{myequation}
where $X$ is either $M$ or $\bar{\lambda}^{tid}$, and rand$\left[A,B\right]$ represents the standard uniform distribution in the interval $[A,B]$. We will consider the same error $\epsilon_M=\epsilon_{\bar{\lambda}^{tid}}=0.01$ for both the mass and the tidal Love parameter.

The objective of this analysis is to find out how an experimental error, i.e. $\Delta M_{\text{exp}}$ and $\Delta \bar{\lambda}^{tid}_{\text{exp}}$, would propagate to the polytropic parameters of the reconstructed equation of state.

We will carry out several realizations of the first iteration of the meshing and refinement method with different inputs. We here present three typical realizations to show the characteristics of the results.

\vspace{0.07cm}\begin{center} \textbf{First realization of the test}\end{center}\vspace{0.07cm}

The results of the first realization of the test are shown in TABLE \ref{tabla6}.

\begin{table}[H]
	\begin{center}
		\begin{tabular}{| c | c | c | c | c |}\hline
			$e_i$ & $\log_{10} p_1$ & $\Gamma_1$ & $\Gamma_2$ & $\Gamma_3$\\\hline
0.048347 & 34.233 & 2.28 & 3.5333 & 3.4\\
0.055149 & 34.187 & 2.14 & 3.8 & 3.1333\\
0.059034 & 34.233 & 4.1 & 3.5333 & 3.2667\\\hline
		\end{tabular}
	\end{center}
	\captionsetup{width=0.5\textwidth}\caption{Polytropic parameters of the first 3 EOSs with the smallest values of $e_i$ in the $4$-volume defined by eq. \eqref{parametrosI}.}\label{tabla6}
\end{table}

\vspace{0.07cm}\begin{center} \textbf{Second realization of the test}\end{center}\vspace{0.07cm}

The results of the second realization of the test are shown in TABLE \ref{tabla7}.

\begin{table}[H]
	\begin{center}
		\begin{tabular}{| c | c | c | c | c |}\hline
			$e_i$ & $\log_{10} p_1$ & $\Gamma_1$ & $\Gamma_2$ & $\Gamma_3$\\\hline
0.056558 & 34.233 & 2 & 3.6667 & 3.2667\\
0.060628 & 34.233 & 2.84 & 3.6667 & 3.1333\\
0.063498 & 34.233 & 2.98 & 3.6667 & 3.1333\\\hline
		\end{tabular}
	\end{center}
	\captionsetup{width=0.5\textwidth}\caption{Polytropic parameters of the first 3 EOSs with the smallest values of $e_i$ in the $4$-volume defined by eq. \eqref{parametrosI}.}\label{tabla7}
\end{table}

\vspace{0.07cm}\begin{center} \textbf{Third realization of the test}\end{center}\vspace{0.07cm}

The results of the third realization of the test are shown in TABLE \ref{tabla8}.

\begin{table}[H]
	\begin{center}
		\begin{tabular}{| c | c | c | c | c |}\hline
			$e_i$ & $\log_{10} p_1$ & $\Gamma_1$ & $\Gamma_2$ & $\Gamma_3$\\\hline
0.066252 & 34.187 & 2 & 3.8 & 3.2667\\
0.069717 & 34.233 & 2.56 & 3.5333 & 3.5333\\
0.073072 & 34.233 & 2.7 & 3.5333 & 3.5333\\\hline
		\end{tabular}
	\end{center}
	\captionsetup{width=0.5\textwidth}\caption{Polytropic parameters of the first 3 EOSs with the smallest values of $e_i$ in the $4$-volume defined by eq. \eqref{parametrosI}.}\label{tabla8}
\end{table}

Comparing the results shown in TABLEs \ref{tabla6}, \ref{tabla7} and \ref{tabla8} with those shown in TABLE \ref{tabla1}, we conclude that $\Gamma_1$ is the most affected polytropic parameter under a small variation of the original input data.

Note that the results presented here are just an estimation since we only show the first iteration of the meshing and refinement method. If several iterations were applied, we would expect the variations of all the polytropic parameters to be smaller.

\section{Restrictions on the piecewise polytropic parameters given by the GW170817 event}\label{constraintsGW170817}

In section \ref{problema_inverso} we generated a mesh of EOSs, given by eq. \eqref{parametrosI}, and calculated the $\bar{\lambda}^{tid}(M)$ curve for each of them ($30$ configurations per EOS). In this section we will check which of these EOSs fulfill the constraints imposed by the GW170817 event.

Consider the waveform model TaylorF2 for the GW170817 event (reference \cite{abbott2019properties}). Let us calculate the mass that both stars would have if $M_1=M_2$ for the chirp mass $\mathcal{M}=1.186M_\odot$, i.e. the central value of eq. \eqref{chirp_mass}. Using the definition of the chirp mass, eq. \eqref{chirp_mass_equation}, one finds that
\begin{myequation}
\mathcal{M}=\frac{(x^2)^{3/5}}{(2x)^{1/5}}=1.186M_\odot \to x\approx 1.3624M_\odot.
\end{myequation}
If $M_1=M_2\approx 1.3624M_\odot$, then $\bar{\lambda}^{tid}_1=\bar{\lambda}^{tid}_2$. Let us define the $\bar{\lambda}^{tid}$ value where the $90\%$ credible region intersects with the $\bar{\lambda}^{tid}_1=\bar{\lambda}^{tid}_2$ curve as $\bar{\lambda}^{tid}_{90\%}$,
\begin{myequation}
\bar{\lambda}^{tid}_{90\%}\approx  842.1.
\end{myequation}
We will say that an EOS lies outside the $90\%$ confidence contour if $\bar{\lambda}^{tid}(M=1.3624M_\odot)>\bar{\lambda}^{tid}_{90\%}$, i.e. for a given EOS we will use the following algorithm:
\begin{itemize}
\item calculate $\bar{\lambda}^{tid}(M)$. In particular, calculate with a certain precision $\bar{\lambda}^{tid}(M=1.3624M_\odot)$. This is only possible if the EOS reaches $1.3624M_\odot$.
\item compare $\bar{\lambda}^{tid}(M=1.3624M_\odot)$ with $\bar{\lambda}^{tid}_{90\%}$.
\item if $\bar{\lambda}^{tid}(M=1.3624M_\odot)\leq\bar{\lambda}^{tid}_{90\%}$, we will say that the EOS lies inside the $90\%$ credible region. Otherwise, we will say that the EOS lies outside the $90\%$ credible region.
\end{itemize}
We will refer to the previous algorithm as the $\bar{\lambda}^{tid}_1=\bar{\lambda}^{tid}_2$ criterion. We will apply the $\bar{\lambda}^{tid}_1=\bar{\lambda}^{tid}_2$ criterion to the mesh of EOSs given by eq. \eqref{parametrosI}, which has a total of $16^4=65536$ EOSs. For each of these EOSs, we have already calculated $30$ stellar configurations. Now we are interested in calculating the $1.3624M_\odot$ configuration with a certain precision, which will be taken as $0.05\%$. 

\begin{figure}[H]
\centering
\includegraphics[width=0.95\linewidth]{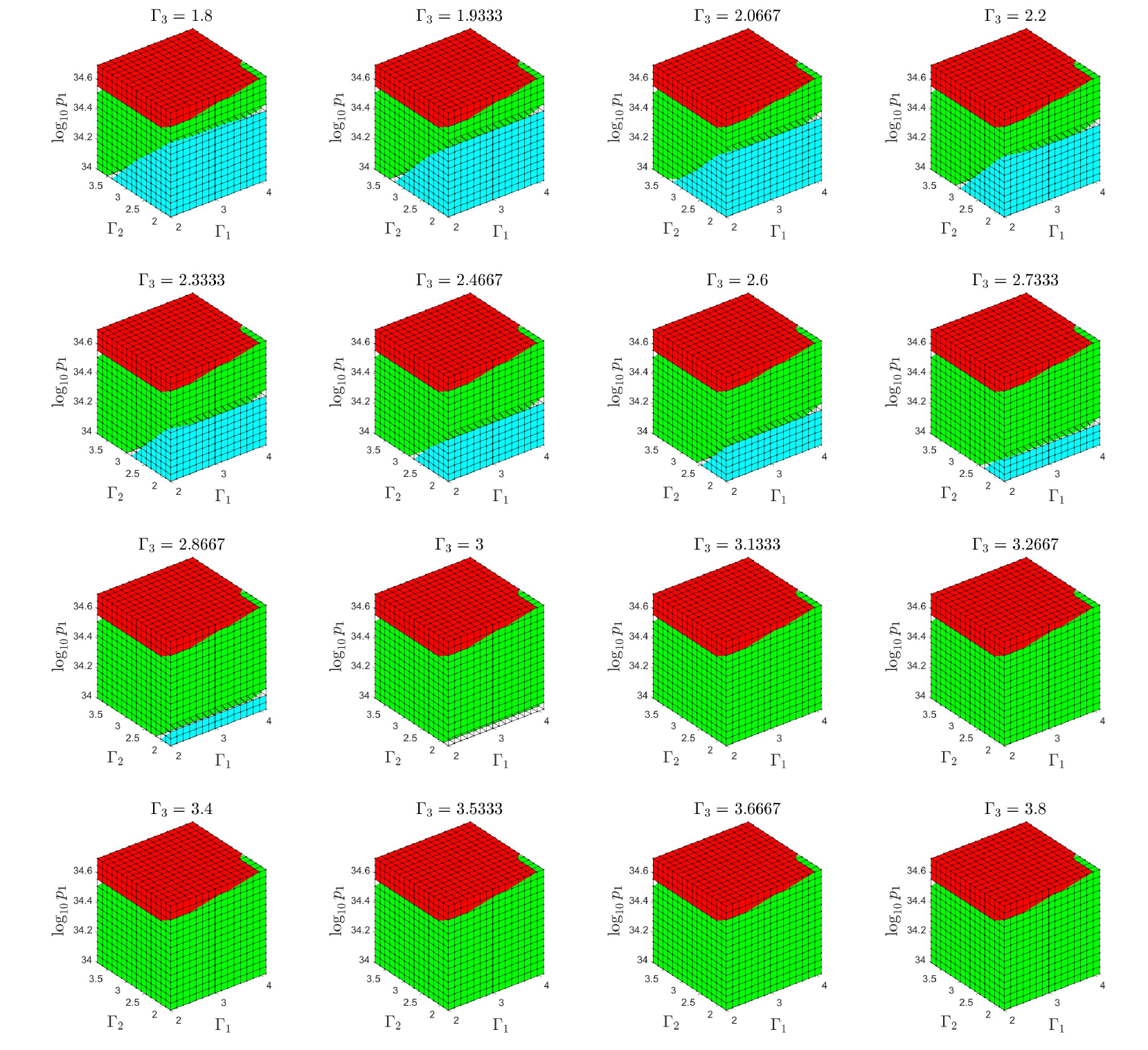}\captionsetup{width=\textwidth}\caption{Constraints on the piecewise polytropic parameters given by the waveform model TaylorF2 for the GW170817 event. The cyan region represents EOSs that do not reach $1.3624M_\odot$. Of the EOSs that reach $1.3624M_\odot$, the green region represents those with $\bar{\lambda}^{tid}(M=1.3624M_\odot)\leq 842.1$, while the red region represents those with $\bar{\lambda}^{tid}(M=1.3624M_\odot)> 842.1$.}\label{fig19}
\end{figure}

Since we cannot represent the four polytropic parameters in a single 4D plot, we will consider the $16$ different $3$-volumes that arise from fixing the value of $\Gamma_3$. This means that each $3$-volume will be given by
\begin{myequation}\label{volumes}
\begin{array}{ll}
\{\log_{10}p_1,\Gamma_1,\Gamma_2,\Gamma_3\}\\\\
=\{[34,34.7]_{16},[2,4.1]_{16},[1.8,3.8]_{16},\text{fixed}\}.
\end{array}
\end{myequation}
The results of applying the $\bar{\lambda}^{tid}_1=\bar{\lambda}^{tid}_2$ criterion to the mesh of EOSs given by eq. \eqref{parametrosI} are shown in FIG. \ref{fig19}, where
\begin{itemize}
\item EOSs that do not reach $1.3624M_\odot$ are represented as the cyan region.
\item EOSs that reach $1.3624M_\odot$:
\begin{itemize}
\item EOSs with $\bar{\lambda}^{tid}_i(M=1.3624M_\odot)\leq \bar{\lambda}^{tid}_{90\%}$ are represented as the green region.
\item EOSs with $\bar{\lambda}^{tid}_i(M=1.3624M_\odot)> \bar{\lambda}^{tid}_{90\%}$ are represented as the red region.
\end{itemize}
\end{itemize}

\onecolumngrid

\vspace{0.4cm}

\twocolumngrid

In all the plots shown in FIG. \ref{fig19}, the excluded red region is the same, i.e. it does not depend on the value of $\Gamma_3$. As one would expect, the excluded cyan region shrinks as $\Gamma_3$ increases, since greater values of $\Gamma_3$ for fixed values of $\{\log_{10}p_1,\Gamma_1,\Gamma_2\}$ mean greater mass values. Because of this, the allowed green region increases in size as $\Gamma_3$ increases.

\subsection{GW170817 event restrictions on piecewise polytropic parameters together with the $2M_\odot$ constraint}

\begin{figure}[H]
\centering
\includegraphics[width=0.98\linewidth]{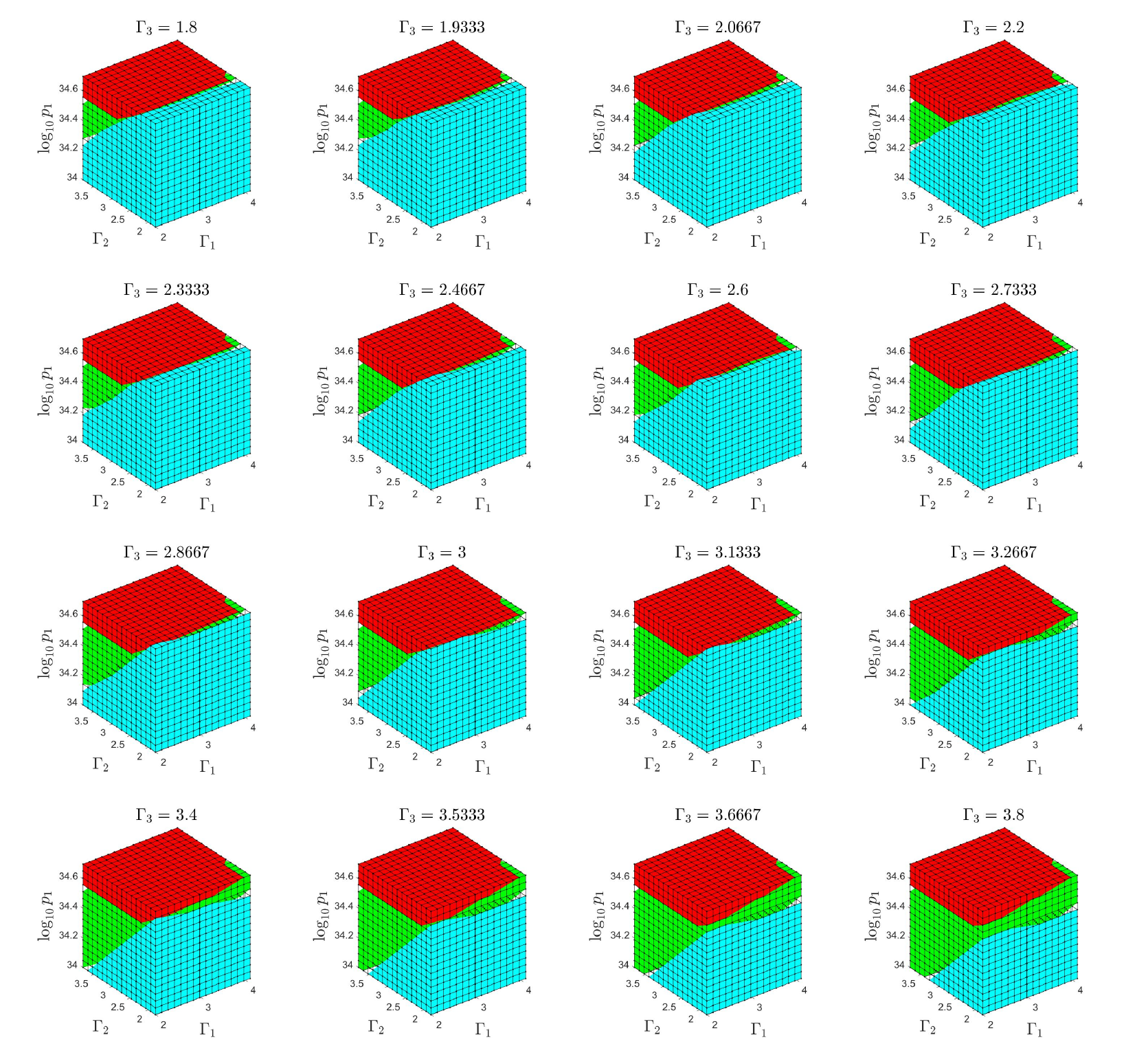}\captionsetup{width=\textwidth}\caption{Constraints on the piecewise polytropic parameters given by the waveform model TaylorF2 for the GW170817 event. The cyan region represents EOSs that do not reach $2M_\odot$. Of the EOSs that reach $2M_\odot$, the green region represents those with $\bar{\lambda}^{tid}(M=1.3624M_\odot)\leq 842.1$, while the red region represents those with $\bar{\lambda}^{tid}(M=1.3624M_\odot)> 842.1$.}\label{fig19_2}
\end{figure}

Here, on apart from constraining the piecewise polytropic parameters $\{\log_{10}p_1,\Gamma_1,\Gamma_2,\Gamma_3\}$ with the $\bar{\lambda}^{tid}_1=\bar{\lambda}^{tid}_2$ criterion, we will also take into account the $2M_\odot$ constraint. The results are shown in FIG. \ref{fig19_2}, where
\begin{itemize}
\item EOSs that do not reach $2M_\odot$ are represented as the cyan region.
\item EOSs that reach $2M_\odot$:
\begin{itemize}
\item EOSs with $\bar{\lambda}^{tid}_i(M=1.3624M_\odot)\leq \bar{\lambda}^{tid}_{90\%}$ are represented as the green region.
\item EOSs with $\bar{\lambda}^{tid}_i(M=1.3624M_\odot)> \bar{\lambda}^{tid}_{90\%}$ are represented as the red region.
\end{itemize}
\end{itemize}

\onecolumngrid

\twocolumngrid

The results shown in FIG. \ref{fig19_2} are similar to the ones shown in FIG. \ref{fig19}, but there are some differences:
\begin{itemize}
\item In FIG. \ref{fig19_2} the cyan region is larger for every value of $\Gamma_3$ than in FIG. \ref{fig19}. This is the expected result since the cyan region now represents the EOSs that do not reach $2M_\odot$ instead of $1.3624M_\odot$. Because of this, the allowed green region is now smaller for every value of $\Gamma_3$.
\item In FIG. \ref{fig19_2}, the excluded red region is suppressed by the cyan one for small values of $\Gamma_3$, which did not happen in FIG. \ref{fig19}.
\end{itemize}

\section{Conclusions}\label{conclusions}

The main objective of this paper was the development of a method to reconstruct the neutron star EOS from measurements of the mass and the tidal Love parameter of different neutron stars. The method is based in the one presented in reference \cite{mena2019reconstruction}. It starts with a wide mesh of polytropic parameters ($65536$ EOSs) which is locally refined in the subsequent iterations. We have tested it considering the input data as $6$ APR4 configurations ($6$ values of $M_{\text{APR4}}$ and $\bar{\lambda}^{tid}_{\text{APR4}}$) and found that the algorithm reconstructs the EOS up to a given tolerance. The reconstructed EOS and the original APR4 are very similar since the polytropic parameters of both EOS are similar itself. Moreover, the macroscopic parameters calculated from the reconstructed EOS are very similar to the ones calculated from the original APR4 EOS. We are confident that the method would work efficiently with experimental data. Also, the algorithm is designed in such a way that it can reconstruct the EOS even if its polytopic parameters do not belong to the initial mesh. We also studied the effect that an experimental error would have on the reconstructed EOSs, and concluded that $\Gamma_1$ is the most affected polytropic parameter under a small variation of the original input data.

As a subproduct of the meshing and refinement method, we have studied which EOSs of the initial mesh of polytropic parameters fulfill the constraints imposed by the GW170817 event. We used the waveform model TaylorF2 for the low-spin scenario, and showed that the EOSs that lie outside the $90\%$ credible region when $\bar{\lambda}^{tid}_1=\bar{\lambda}^{tid}_2$ define a zone of polytropic parameters that does not depend on $\Gamma_3$. We also showed that the excluded region with the $2M_\odot$ constraint decreases as $\Gamma_3$ increases, which gives raise to larger allowed regions for greater values of $\Gamma_3$.

\section{Acknowledgments}

We would like to thank J. L. Blázquez Salcedo and F. Navarro Lérida for their comments, suggestions and very helpful discussions.

\onecolumngrid

\bibliographystyle{unsrt}
\bibliography{inverse_love}

\begin{thebibliography}{10}

\bibitem{abbott2016observation}
Benjamin~P Abbott, Richard Abbott, TD~Abbott, MR~Abernathy, Fausto Acernese,
  Kendall Ackley, Carl Adams, Thomas Adams, Paolo Addesso, RX~Adhikari, et~al.
\newblock Observation of gravitational waves from a binary black hole merger.
\newblock {\em Physical review letters}, 116(6):061102, 2016.

\bibitem{abbott2017gw170814}
Benjamin~P Abbott, R~Abbott, TD~Abbott, F~Acernese, K~Ackley, C~Adams, T~Adams,
  P~Addesso, RX~Adhikari, VB~Adya, et~al.
\newblock Gw170814: a three-detector observation of gravitational waves from a
  binary black hole coalescence.
\newblock {\em Physical review letters}, 119(14):141101, 2017.

\bibitem{abbott2017gw170817}
Benjamin~P Abbott, Rich Abbott, TD~Abbott, Fausto Acernese, Kendall Ackley,
  Carl Adams, Thomas Adams, Paolo Addesso, RX~Adhikari, VB~Adya, et~al.
\newblock Gw170817: {O}bservation of gravitational waves from a binary neutron
  star inspiral.
\newblock {\em Physical Review Letters}, 119(16):161101, 2017.

\bibitem{lindblom1992determining}
Lee Lindblom.
\newblock Determining the nuclear equation of state from neutron-star masses
  and radii.
\newblock {\em The Astrophysical Journal}, 398:569--573, 1992.

\bibitem{lindblom2012spectral}
Lee Lindblom and Nathaniel~M Indik.
\newblock Spectral approach to the relativistic inverse stellar structure
  problem.
\newblock {\em Physical Review D}, 86(8):084003, 2012.

\bibitem{lindblom2014spectral}
Lee Lindblom and Nathaniel~M Indik.
\newblock Spectral approach to the relativistic inverse stellar structure
  problem {II}.
\newblock {\em Physical Review D}, 89(6):064003, 2014.

\bibitem{lindblom2014relativistic}
Lee Lindblom.
\newblock The relativistic inverse stellar structure problem.
\newblock In {\em AIP Conference Proceedings}, volume 1577, pages 153--164.
  AIP, 2014.

\bibitem{kokkotas2001inverse}
KD~Kokkotas, TA~Apostolatos, and N~Andersson.
\newblock The inverse problem for pulsating neutron stars: a ‘fingerprint
  analysis’ for the supranuclear equation of state.
\newblock {\em Monthly Notices of the Royal Astronomical Society},
  320(3):307--315, 2001.

\bibitem{lackey2015reconstructing}
Benjamin~D Lackey and Leslie Wade.
\newblock Reconstructing the neutron-star equation of state with
  gravitational-wave detectors from a realistic population of inspiralling
  binary neutron stars.
\newblock {\em Physical Review D}, 91(4):043002, 2015.

\bibitem{carney2018comparing}
Matthew~F Carney, Leslie~E Wade, and Burke~S Irwin.
\newblock Comparing two models for measuring the neutron star equation of state
  from gravitational-wave signals.
\newblock {\em Physical Review D}, 98(6):063004, 2018.

\bibitem{abdelsalhin2018solving}
Tiziano Abdelsalhin, Andrea Maselli, and Valeria Ferrari.
\newblock Solving the relativistic inverse stellar problem through
  gravitational waves observation of binary neutron stars.
\newblock {\em Physical Review D}, 97(8):084014, 2018.

\bibitem{abbott2018gw170817}
BP~Abbott, R~Abbott, TD~Abbott, F~Acernese, K~Ackley, C~Adams, T~Adams,
  P~Addesso, RX~Adhikari, VB~Adya, et~al.
\newblock Gw170817: {M}easurements of neutron star radii and equation of state.
\newblock {\em Physical review letters}, 121(16):161101, 2018.

\bibitem{volkel2019inverse}
Sebastian~H V{\"o}lkel and Kostas~D Kokkotas.
\newblock On the {I}nverse {S}pectrum {P}roblem of {N}eutron {S}tars.
\newblock {\em arXiv preprint arXiv:1901.11262}, 2019.

\bibitem{raithel2018tidal}
Carolyn~A Raithel, Feryal {\"O}zel, and Dimitrios Psaltis.
\newblock Tidal deformability from {GW}170817 as a direct probe of the neutron
  star radius.
\newblock {\em The Astrophysical Journal Letters}, 857(2):L23, 2018.

\bibitem{chatziioannou2018measuring}
Katerina Chatziioannou, Carl-Johan Haster, and Aaron Zimmerman.
\newblock Measuring the neutron star tidal deformability with
  equation-of-state-independent relations and gravitational waves.
\newblock {\em Physical Review D}, 97(10):104036, 2018.

\bibitem{most2018new}
Elias~R Most, Lukas~R Weih, Luciano Rezzolla, and J{\"u}rgen Schaffner-Bielich.
\newblock New constraints on radii and tidal deformabilities of neutron stars
  from {GW}170817.
\newblock {\em Physical review letters}, 120(26):261103, 2018.

\bibitem{de2018tidal}
Soumi De, Daniel Finstad, James~M Lattimer, Duncan~A Brown, Edo Berger, and
  Christopher~M Biwer.
\newblock Tidal {D}eformabilities and {R}adii of {N}eutron {S}tars from the
  {O}bservation of {GW}170817.
\newblock {\em Physical review letters}, 121(9):091102, 2018.

\bibitem{mena2019reconstruction}
Juan Mena-Fern{\'a}ndez and Luis~Manuel Gonz{\'a}lez-Romero.
\newblock Reconstruction of the neutron star equation of state from
  w-quasinormal modes spectra with a piecewise polytropic meshing and
  refinement method.
\newblock {\em arXiv preprint arXiv:1901.10851}, 2019.

\bibitem{regge1957stability}
Tullio Regge and John~A Wheeler.
\newblock Stability of a {S}chwarzschild singularity.
\newblock {\em Physical Review}, 108(4):1063, 1957.

\bibitem{thorne1967non}
Kip~S Thorne and Alfonso Campolattaro.
\newblock Non-radial pulsation of general-relativistic stellar models. {I}.
  {A}nalytic analysis for l>= 2.
\newblock {\em The astrophysical journal}, 149:591, 1967.

\bibitem{hinderer2008tidal}
Tanja Hinderer.
\newblock Tidal {L}ove numbers of neutron stars.
\newblock {\em The Astrophysical Journal}, 677(2):1216, 2008.

\bibitem{hinderer2010tidal}
Tanja Hinderer, Benjamin~D Lackey, Ryan~N Lang, and Jocelyn~S Read.
\newblock Tidal deformability of neutron stars with realistic equations of
  state and their gravitational wave signatures in binary inspiral.
\newblock {\em Physical Review D}, 81(12):123016, 2010.

\bibitem{akmal1998equation}
A~Akmal, VR~Pandharipande, and DG~Ravenhall.
\newblock Equation of state of nucleon matter and neutron star structure.
\newblock {\em Physical Review C}, 58(3):1804, 1998.

\bibitem{douchin2001unified}
F~Douchin and P~Haensel.
\newblock A unified equation of state of dense matter and neutron star
  structure.
\newblock {\em Astronomy \& Astrophysics}, 380(1):151--167, 2001.

\bibitem{glendenning1984neutron}
Norman~K Glendenning.
\newblock Neutron stars are giant hypernuclei?
\newblock 1984.

\bibitem{bednarek2012hyperons}
I~Bednarek, P~Haensel, JL~Zdunik, M~Bejger, and R~Ma{\'n}ka.
\newblock Hyperons in neutron-star cores and a two-solar-mass pulsar.
\newblock {\em Astronomy \& Astrophysics}, 543:A157, 2012.

\bibitem{alford2005hybrid}
Mark Alford, Matt Braby, Mark Paris, and Sanjay Reddy.
\newblock Hybrid stars that masquerade as neutron stars.
\newblock {\em The Astrophysical Journal}, 629(2):969, 2005.

\bibitem{bonanno2012composition}
Luca Bonanno and Armen Sedrakian.
\newblock Composition and stability of hybrid stars with hyperons and quark
  color-superconductivity.
\newblock {\em Astronomy \& Astrophysics}, 539:A16, 2012.

\bibitem{weissenborn2011quark}
Simon Weissenborn, Irina Sagert, Giuseppe Pagliara, Matthias Hempel, and
  J{\"u}rgen Schaffner-Bielich.
\newblock Quark matter in massive compact stars.
\newblock {\em The Astrophysical Journal Letters}, 740(1):L14, 2011.

\bibitem{abbott2019properties}
BP~Abbott, R~Abbott, TD~Abbott, F~Acernese, K~Ackley, C~Adams, T~Adams,
  P~Addesso, RX~Adhikari, VB~Adya, et~al.
\newblock Properties of the binary neutron star merger {GW}170817.
\newblock {\em Physical Review X}, 9(1):011001, 2019.

\bibitem{tauris2017formation}
TM~Tauris, M~Kramer, PCC Freire, N~Wex, H-T Janka, N~Langer, Ph~Podsiadlowski,
  E~Bozzo, S~Chaty, MU~Kruckow, et~al.
\newblock Formation of double neutron star systems.
\newblock {\em The Astrophysical Journal}, 846(2):170, 2017.

\bibitem{read2009constraints}
Jocelyn~S Read, Benjamin~D Lackey, Benjamin~J Owen, and John~L Friedman.
\newblock Constraints on a phenomenologically parametrized neutron-star
  equation of state.
\newblock {\em Physical Review D}, 79(12):124032, 2009.

\end{thebibliography}

\end{document}